\documentclass[twocolumn, aps, prd, superscriptaddress,showpacs, floatfix]{revtex4}

\usepackage{multirow}
\usepackage{epsfig}
\usepackage{amsmath}

\usepackage{latexsym}
\usepackage{feynmp}

\usepackage[normalem]{ulem}  
\usepackage[dvips]{color} 

\renewcommand\sout{\bgroup \color{red} \ULdepth=-.5ex \ULset}

\begin{document}

\title{A novel probe of chiral restoration in nuclear medium}

\author{Philipp Gubler}
\affiliation{Department of Physics and Institute of Physics and Applied Physics, Yonsei
University, Seoul 120-749, Korea}
\author{Teiji Kunihiro}
\affiliation{Department of Physics, Faculty of Science, Kyoto University, Kyoto 606-8502, Japan}
\author{Su Houng Lee}
\affiliation{Department of Physics and Institute of Physics and Applied Physics, Yonsei
University, Seoul 120-749, Korea}

\begin{abstract}
We propose measuring the mass shift and width broadening of the $f_1(1285)$ meson 
 together with those of the $\omega$ from a nuclear target as a  means 
to  experimentally probe the partial restoration of chiral symmetry inside the nuclear matter. 
The relation between the order parameter of chiral symmetry and the difference in the correlation functions of the $f_1(1285)$ current and the $\omega$ current  is discussed 
in the limit where the disconnected diagrams are neglected.
A QCD sum rule analysis of the $f_1(1285)$ meson mass leads to about 100 MeV attraction in nuclear matter, 
which can be probed in future experiments.  
\end{abstract}



\maketitle

\section{Introduction}

A long standing and prominent problem in nuclear and hadron physics is understanding the origin of the hadron masses, 
that comprise the dominant part of the mass of our visible universe 
\cite{Hatsuda:1985eb,Brown:1991kk,Hatsuda:1991ez,Leupold:2009kz}.   
While chiral symmetry breaking is expected to be responsible for generating hadron masses of the order 
of GeV starting from the bare quark masses that are smaller than 10 MeV, 
it is not clear how the effect is manifested inside a hadron, where confinement makes the quark 
an unobservable gauge dependent object.  

There is a well defined  order parameter of chiral symmetry breaking: 
the chiral condensate $\langle \bar{q} q \rangle $.  However it is a challenge to relate it to physically 
accessible quantities in a model independent way \cite{Jido:2008bk}.  
The differences between current-current correlation functions of chiral partners are considered to be another promising set of order parameters of chiral symmetry.  
In fact, it was shown that the density of zero eigenvalues of the Dirac equation, responsible for generating 
the chiral order parameter \cite{Banks:1979yr}, is also responsible for breaking the degeneracies 
of the corresponding correlation functions \cite{Cohen:1996ng} apart from the $U_A(1)$ breaking 
effect coming from contributions of  topologically non trivial configurations \cite{Lee:1996zy}.  

While the whole correlation function should be measured to verify the restoration of the chiral symmetry breaking, 
the ground state poles that couple to the correlation functions are the most distinctive feature 
of the correlation functions, and hence measuring the mass differences between chiral partners is considered 
to be the most attractive alternative.  Moreover, chiral symmetry is expected to be restored in the spectrum 
of excited states\cite{Glozman:2007ek}.    
 The first works on the temperature dependence of the sigma-pion mass \cite{Hatsuda:1985eb,Hatsuda:1986gu} showed that 
the masses of the chiral partners become  degenerate  near the chiral phase transition point.  
The vector- axialvector masses \cite{Hatsuda:1992bv} were also found to decrease at finite temperature.   
One could think of measuring such mass shift of hadrons in heavy ion collisions.  
However, whatever signal there is near the phase transition, it will be lost during the hadronic 
evolution of the system.  In fact, signals appearing as peaks in the low mass dilepton spectrum 
were found to be dominated by a large broadening of the $\rho$ meson \cite{Rapp:1997fs}.

Measuring the mass shift from the nuclear target was suggested as an alternative because the chiral order 
parameter can be shown to be quenched by more than 30\% in nuclear medium. Moreover, the nuclear target would 
provide a stable environment where the density profile is fixed so that the effects of time  
evolution can be neglected \cite{Leupold:2009kz}.  
Unfortunately, so far, most attempts were problematic because the width broadening of the already wide mesons 
made any realistic measurement not meaningful.  However, the CBELSA/TAPS collaboration have been 
focusing on the $\omega$ and $\eta'$ meson and successfully measured the mass and with broadening 
of these particles \cite{Metag:2015lza,Nanova:2016cyn}. The main reason behind their success is 
in their focus on mesons with small vacuum width so that a non trivial proportional 
increase in the meson's width will still leave the meson narrow enough for a meaningful measurement 
to identify and describe the meson in the nuclear medium. 

Recently, the CLAS collaboration has successfully identified the $f_1(1285)$ in photoproduction 
from a proton target with a small width of $18 \pm1.4$ MeV \cite{Dickson:2016gwc}. 
As we will discuss in the next section, the $f_1(1285)$ can be considered to be the chiral partner of the $\omega$ 
if disconnected diagrams are neglected.
Hence, when the experiment is applied to a nuclear target and when combined with observations 
for the $\omega$ meson, one can finally hope to observe and experimentally verify the partial 
chiral symmetry restoration expected to occur in nuclear medium. 
Such an observation will provide critical information on the origin of hadron masses.

\section{Ideal mixing in the vector and axial vector channels}

When we restrict ourselves to two flavors, the chiral partners of $SU(2)_L \times SU(2)_R$ 
in the vector-axial vector channel are the $\rho$ and $a_1$ mesons.  Both have large width even in vacuum 
and are hence not good candidates to be measured in nuclear medium. Extending the flavor number to three, 
one finds that the octet meson is  mixed with the singlet, forming a nonet.  
For the vector channel the extra mesons with isospin zero mixes with the octet almost ideally.  
This means that one  can identify the quark content as $\omega= \frac{1}{2} (\bar{u} u+ \bar{d}d)$ and $\phi=\bar{s}s$.
At the same time, the $\omega$ mass is almost degenerate with that of the $\rho$, which comes about naturally 
in the large $N_c$ limit where we can neglect the correlation function between $u$-quark current and the $d$ quark current.

In fact, ideal mixing in the three flavor case and the suppression of disconnected diagrams are related. To see
this we recapitulate here briefly why we have ideal mixing in the vector channel, where the suppression of disconnected
diagrams is largest. To visualize, it is convenient to just consider the two point function composed of the singlet
$\omega_1 = \frac{1}{\sqrt{3}} 
(\overline{u}u + \overline{d}d + \overline{s}s)$ and the octet $\omega_8 = 
\frac{1}{\sqrt{6}}(\overline{u}u + \overline{d}d - 2\overline{s}s)$ currents. 
The correlation functions can be represented
as a two by two matrix composed of the elements $\Pi_{ij} = \langle \omega_i, \omega_j \rangle$. 
If we neglect the disconnected diagrams and work in the quenched approximation, this matrix can be written as follows
\begin{equation}
\begin{pmatrix} 
\Pi_0 + \frac{1}{3} \Delta \Pi_s & - \frac{\sqrt{2}}{3} \Delta \Pi_s \\
- \frac{\sqrt{2}}{3} \Delta \Pi_s & \Pi_0 + \frac{2}{3} \Delta \Pi_s 
\end{pmatrix},
\end{equation}
where 
\begin{eqnarray}
\Pi_0 &=&  \langle \omega_1, \omega_1 \rangle_{m_s = 0}, \\
\Delta \Pi_s &=& \langle \overline{s}s,   \overline{s}s\rangle - 
\langle \overline{s}s,   \overline{s}s\rangle_{m_s = 0}, 
\end{eqnarray}
That is, $\Pi_0$ represents the SU(3) symmetric correlation function, while the $\Delta \Pi_s$ encodes the symmetry breaking part
and the mixing between the singlet and octet. Diagonalizing, we find the ideally mixed correlation. If we include the
disconnected diagram, we will not obtain ideal mixing. Therefore, whether we are in two or three flavors, 
when disconnected diagrams are neglected, the $\omega$ becomes degenerate with the $\rho$.

In the axial vector channel, the mixing angle between the two isospin zero mesons $f_1(1285)$, $f_1(1420)$ 
is not determined as well as in the case of the vector mesons. 
Some hadronic models in fact obtain a rather large strangeness component of the $f_1(1285)$ \cite{Lutz:2003fm}.
Most estimates however find that the quark content of the $f_1(1285)$ is dominated by 
$u$ and $d$ quarks with only a small $s$ component \cite{Close:2015rza,Neumeier:2000fb,Li:2000dy}. 
Our own analysis based on QCD sum rules, to be described in the next section, points to the same conclusion.
Moreover, one should note that the mass of $f_1(1285)$ is almost degenerate with that 
of the $a_1$, as the $\omega$ is with the $\rho$. 
All this suggests that the $f_1(1285)$ and $f_1(1420)$ are almost ideally mixed, demonstrating that the 
disconnected diagrams are also suppressed in this channel. 
Within this limit, one can argue that the $\omega$ and the $f_1(1285)$ are chiral partners and will become
degenerate when the chiral symmetry gets restored, which was also anticipated in Ref~\cite{Rapp:1999ej}. 

Let us discuss this point in some more detail. In other words:  
why can $\omega$ and $f_1(1285)$ be considered as chiral partners when disconnected diagrams are neglected, 
while they do not seem to be directly related by chiral SU(2) symmetry? 
This is related to why the huge $\omega$ or  $f_1(1285)$ masses, relative to the current quark masses, are almost degenerate to that of the $\rho$ and $a_1$ respectively. 
To understand this, one notes that using the Casher Banks formula, one
finds that spontaneous chiral symmetry breaking occurs when the density of zero modes of the Dirac equation within the
QCD Euclidean functional integral becomes non zero. Once this density of zero modes becomes non zero, one can show
that all the order parameters of chiral symmetry breaking become non zero. One can furthermore show that when one
neglects the disconnected diagrams, the difference between the correlation functions of $\omega$ and $f_1(1285)$ is proportional
to the density of zero modes \cite{Cohen:1996ng,Lee:1996zy,Lee:2013es}, 
and are in fact identical to that of $\rho$ and $a_1$. 
Now, even if one includes the disconnected diagrams, one notes that their contributions 
are small in the vacuum   as the leading contribution requires at least three gluon exchange.  
Moreover, the effect from the $U_A(1)$ anomaly is small in the vacuum. This is so because the phenomenologically observable effect of the $U_A(1)$ 
anomaly comes in through topologically non-trivial configurations that connects zero modes of different chirality. Such configurations
have minimal effects in the vector or axial channel. One can visualize such effects as an instanton contribution that has
left handed and right handed quarks with $N_F$ flavors interpolating between correlation functions of the two currents; if
the currents are vector or axial vector currents, a single instanton can not interpolate the correlation function as left
right symmetry is conserved in such currents. 

In nuclear matter, the non-strange chiral condensate $\langle \overline{u} u + \overline{d} d\rangle$ 
is expected to be partially quenched. From a theoretical point of view, this originates from 
the identical reduction of the zero-modes of the the $u$ and $d$ quarks. 
On the other hand, if one still neglects disconnected diagrams, ideal mixing of $\omega$ and 
$f_1(1285)$ will not be modified even in nuclear matter as 
any additional correlations between strange and non-strange components 
vanish in this limit. This can be understood from a large $N_c$ argument, as in the large $N_c$ limit the 
nucleon has no sea quarks and therefore no strangeness content. 
Next, considering only $u$ and $d$ quarks and assuming that the correlation between disconnected quark lines within a nucleon is suppressed compared to 
those that are connected, one concludes that 
the changes of the zero modes will affect the difference between the $\omega$ and 
the $f_1(1285)$ in the same way as it does for the $\rho$ and $a_1$, when disconnected diagrams are neglected. 
Any residual $U_A(1)$ anomaly breaking effect in the presence of nucleons can in principle be estimated
separately by measuring the $\eta'$ in nuclear matter, which has been made possible by the experimental efforts reported in
Ref.\,\cite{Nanova:2016cyn}. 
Therefore when the measurements of all the small width mesons, $\eta'$, $\omega$ and $f_1(1285)$ are combined, we will 
finally have a better understanding of how, if at all, chiral symmetry and $U_A(1)$ are partially restored in nuclear medium, and whether they are at all responsible for generating hadron masses.

\begin{table}
\caption{ Width and Mass of chiral partners. Units are in MeV.}
\begin{center}
\begin{tabular}{|c|c|c||c|c|c|}
\hline \hline
$J^{PC}=1^{--}$    & mass  & width  & $J^{PC}=1^{++}$ & mass & width \\
\hline
$\rho$ & 770 & 150 & $a_1$ & 1260 & 250-600  \\
\hline
$\omega$ & 782 & 8.49 & $f_1$ & 1285 & 24.2  \\
\hline
$\phi$ & 1020 & 4.266 & $f_1$ & 1420 & 54.9  \\
\hline \hline
\end{tabular}
\end{center}
\label{mass-width}
\end{table}

\section{QCD sum rule analysis}

We first give a quick overview of the QCD sum rule analysis method adapted in this 
work. 

The starting point is the correlation function of the axial vector current in the nuclear medium, 
\begin{eqnarray}
\Pi_{\mu \nu} (\omega, {\bf q}) & = & i \int d^4x e^{iqx} \langle {\rm T} J_\mu(x) J_\nu(0) \rangle_{n.m.},
\end{eqnarray}
where the subscript $n.m.$ stands for the nuclear matter expectation value and $q_\mu=(\omega,{\bf q})$.  
The current is taken to be either 
$J^q_\mu=\eta_{\mu \nu} \frac{1}{\sqrt{2}} \langle \bar{u} \gamma_\nu \gamma_5 u+ \bar{d} \gamma_\nu \gamma_5 d \rangle$ or 
$J^s_\mu=\eta_{\mu \nu} \langle \bar{s} \gamma_\nu \gamma_5 s \rangle$,  
where $\eta_{\mu \nu}=q_\mu q_\nu/q^2 - g_{\mu \nu}$.  
We will closely follow the finite temperature formalism given in Ref.~\cite{Hatsuda:1992bv} 
and apply it to finite density \cite{Hatsuda:1991ez}. We look at the trace part of the polarization function 
$\Pi(Q^2)= -\Pi^\mu_\mu(Q^2)/3$ at ${\bf q} \rightarrow 0$.  The OPE for $J^q_\mu$ is given as 
\begin{eqnarray}
&& \Pi(Q^2) = \nonumber \\ 
&& \frac{1}{4 \pi^2} \bigg(1+\frac{\alpha_s}{\pi} \bigg) Q^2\ln ( Q^2) 
+ \frac{3}{2 \pi^2} m_q^2 \ln ( Q^2) \nonumber \\
&& + 2 \frac{m_q}{Q^2} \langle \bar{q} q \rangle 
- \frac{1}{12Q^2} \langle \frac{\alpha_s}{\pi}G^2 \rangle  
+ \frac{2\pi \alpha_s}{Q^4} \langle (\bar{q}\gamma_\mu \lambda^a q)^2 \rangle \nonumber \\  
&& + \frac{4\pi \alpha_s}{9Q^4} 
\langle (\bar{q}\gamma_\mu \lambda^a q)(\sum_{q}^{u,d,s}\bar{q}\gamma_\mu \lambda^a q)  \rangle  \nonumber \\
&& +\frac{8iq^\mu q^\nu}{3Q^4} \langle (\bar{q} \gamma_\mu D_\nu q)_{ST} \rangle \nonumber \\
&& -\frac{32 i q^\mu q^\nu q^\sigma q^\sigma}{3Q^{8}} \langle (\bar{q} \gamma_\mu D_\nu D_\lambda D_\sigma q)_{ST} \rangle. 
\label{eq:OPE}
\end{eqnarray}
Here, we take the quark operators $\langle \bar{q} ..q \rangle$ to be the average of $u,d$ quark contributions, 
except for the one with the summation sign. 
To get the respective expression for the $J^s_\mu$ correlator, one simply replaces $\langle \bar{q} ..q \rangle$ by 
$\langle \bar{s} ..s \rangle$, again with the exception of the operator behind the summation sign. 
While for $u$ and $d$ quarks, the $m_q^2$ term in Eq.\,(\ref{eq:OPE}) is negligible and can be safely neglected, 
we keep it for the strange quark case.
We have also neglected the twist-4 operators in Eq.\,(\ref{eq:OPE}), 
whose contributions are expected to be small \cite{Hatsuda:1995dy}.

We now follow the standard procedure and use the dispersion relation to relate the OPE  to the spectral density, 
generally consisting of poles and continuum, 
which we represent by a delta function for the lowest pole  
and a step function starting from a continuum threshold $s_0$, respectively; in the present simplest treatment, 
the width of the pole is neglected. 
Performing the Borel transformation and taking the ratio with its derivative, 
one obtains a relation for the mass in terms of the Borel mass: 
\begin{eqnarray}
&& \frac{m_{f_1}^2}{M^2} = \nonumber \\ 
&& \bigg[2\bigg(1+\frac{\alpha_s}{\pi} \bigg) E_2(s_0/M^2) - \frac{a}{M^2} E_1(s_0/M^2) \nonumber \\
&&+ \frac{2(e-f)}{M^6} \bigg] \nonumber \\
&  \times &
\bigg[\bigg(1+\frac{\alpha_s}{\pi} \bigg) E_1(s_0/M^2) - \frac{a}{M^2} E_0(s_0/M^2)  \nonumber \\ 
&&- \frac{-b+c+d}{M^4} -\frac{2(e-f)}{M^6} \bigg]^{-1}. 
\label{mass-borel}
\end{eqnarray}
In the present sum rule, there is no contribution from the nucleon scattering term \cite{Hatsuda:1995dy}.  This is so because the scattering term contributes as $s\delta(s)$ in the imaginary part of $\Pi(Q^2)$.  If we had studied the sum rule for $\Pi(Q^2)/Q^2$, the scattering term would have appeared not only for the $f_1$, but also in the $\omega$, $\rho$ and $a_1$ sum rules with different coefficients.  To linear order in density, additional scattering terms coming from excited nucleon intermediate states  could however be added to the imaginary part.  
Such terms will not be of the delta function type but appear near the excitation energies for the intermediate states and will depend on the quantum numbers of the current.  We leave such a detailed modeling as a future work.  
Here we follow the simple pole ansatz to  estimate the maximum possible mass shift for the  $f_1$ meson and compare it to the vacuum width which is an important criterion to  asses the observability in an actual experiment.
The parameters $a$-$f$ read  
\begin{eqnarray}
a &=& 6 m_q^2, \\
b &=&  4 \pi^2 m_q \langle \bar{q} q \rangle_{\rho} = 4 \pi^2 ( m_q \langle \bar{q} q \rangle_{0} + \sigma_{\pi N}\rho ), \\
c &=& \frac{\pi^2}{3} \langle \frac{\alpha_s}{\pi}G^2 \rangle_{\rho} \\
&=&  \frac{\pi^2}{3} \Bigg( \langle \frac{\alpha_s}{\pi}G^2 \rangle_{0} - \frac{8}{9}(M_N - \sigma_{\pi N} - \sigma_{sN})\rho \Bigg), \\
d &=& 4 \pi^2 M_N A_2^q \rho, \\
e &=& \frac{704 \pi^3 \alpha_s}{4\times 81} \langle \bar{q} q \rangle^2_{\rho}, \\
f &=& \frac{10 \pi^2}{3} M_N^3 A_4^q \rho, 
\end{eqnarray}
and the functions $E_i(s_0/M^2)$ are defined as 
\begin{eqnarray}
E_0(s_0/M^2) &=& 1 -  e^{-s_0/M^2}, \\
E_1(s_0/M^2) &=& 1 - (1 + \frac{s_0}{M^2}) e^{-s_0/M^2}, \\
E_2(s_0/M^2) &=& 1 - (1 + \frac{s_0}{M^2} + \frac{s_0^2}{2M^4}) e^{-s_0/M^2}.
\end{eqnarray}

The threshold parameter is determined by requiring the mass in Eq.~(\ref{mass-borel}) to be most stable within the Borel 
window. The minimum Borel mass is determined  by requiring the contribution coming from the highest OPE 
to be less than 10\% of the whole OPE appearing in the denominator of Eq.~(\ref{mass-borel}).  
The maximum Borel mass is determined by the pole dominance criterion, demanding that the pole contribution 
to the sum rule is larger than that of the continuum.

\subsection{Vacuum analysis of $f_1(1285)$ and $f_1(1420)$}
At first, let us study the masses extracted from the $J^q_{\mu}$ and $J^s_{\mu}$ 
correlators to check whether they can approximately reproduce the energy levels of the physical 
$f_1(1285)$ and $f_1(1420)$ states. If $J^{q}_{\mu}$ ($J^{s}_{\mu}$) couples dominantly to $f_1(1285)$ ($f_1(1420)$), 
this would be a strong indication that these two states are close to ideally mixed. 

\begin{figure}
\begin{center}
\vspace{-0.9cm}
\includegraphics[width=0.50\textwidth]{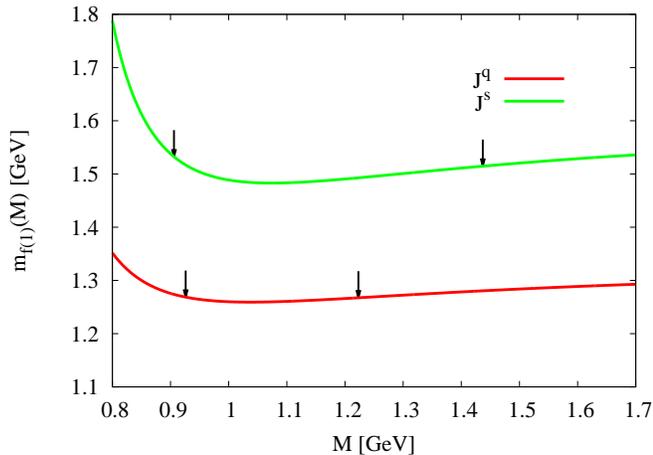}
\vspace{0.3cm}
\end{center}
\caption{The Borel curve for the vacuum mass formula of Eq.~(\ref{mass}) 
for the $J^q_{\mu}$ and $J^s_{\mu}$ correlators. 
The arrows indicate the positions of the respective minimum and maximum Borel masses.}
\label{borel.q.s}
\end{figure}

The Borel curves for the vacuum masses of the $J^q_{\mu}$ and $J^s_{\mu}$ 
correlators are shown in Fig.~\ref{borel.q.s}. 
The parameter values employed to draw these curves are given in the 
upper part of Table~\ref{tab:parameters}. 
\begin{table}
\renewcommand{\arraystretch}{1.5}
\setlength{\tabcolsep}{10pt}
\begin{center}
\caption{Parameter values used in the present calculation. All values were converted to a 
renormalization scale of 1 GeV. The parameters $A^q_2$ and $A^q_4$ are obtained by numerically 
integrating the NLO parton distributions provided in \cite{Martin:2009iq}.} 
\label{tab:parameters}
\begin{tabular}{lc}  
\hline 
$\langle \bar{q} q \rangle_0$ & $(-0.248\,\mathrm{GeV})^3$ \cite{Aoki:2016frl} \\
$\langle \bar{s} s \rangle_0$ & $0.8 \times \langle \bar{q} q \rangle_0$ \cite{Reinders:1984sr} \\
$m_q$ & $4.7\,\mathrm{MeV}$ \cite{Agashe:2014kda} \\
$m_s$ & $95\,\mathrm{MeV}$ \cite{Agashe:2014kda} \\
$\langle \frac{\alpha_s}{\pi}G^2 \rangle_{0}$ & $0.012\,\mathrm{GeV}^4$ \cite{Colangelo:2000dp} \\
$M_N$ & $939\,\,\mathrm{MeV}$ \\
$\sigma_{\pi N}$ & $45\pm15\,\,\mathrm{MeV}$ \cite{Gasser:1990ce,Durr:2015dna,Hoferichter:2015dsa} \\
$\sigma_{sN}$ & $35\,\,\mathrm{MeV}$ \cite{Bali:2016lvx} \\
$A^q_2$ & $0.62$ \cite{Martin:2009iq} \\
$A^q_4$ & $0.066$ \cite{Martin:2009iq} \\
\hline
\end{tabular}
\end{center}
\end{table}
The obtained curves are stable and exhibit a wide Borel window, which indicates that the sum 
rule analysis works well for both cases. The flattest Borel curves give threshold parameters of 
$\sqrt{s_0} = 1.61\,\mathrm{GeV}$ for $J^q_{\mu}$ and 
$\sqrt{s_0} = 1.925\,\mathrm{GeV}$ for $J^s_{\mu}$. 
Comparing the calculated masses with the experimental values, it is seen 
that the result from the $J^q_{\mu}$ correlator agrees almost perfectly with 
the mass of the $f_1(1285)$. The curve extracted from $J^s_{\mu}$ similarly 
lies close to the $f_1(1420)$ mass, however turns out to be about 
$70\,\mathrm{MeV}$ too high. 
Given the slight discrepancy existing in the $f_1(1420)$ sum rule, further theoretical studies are required to shed more light on the possible 
importance of disconnected diagrams in the $f_1$ channel and substantiate the chiral partner scenario. 
Furthermore, there could still be a significant four-quark component  
in the wave functions for both $f_1(1285)$ and $f_1(1420)$, as has been shown for instance in the coupled channel type analysis of Ref.\,\cite{Roca:2005nm}, 
which contributes to both the connected and disconnected diagrams.
Nevertheless, the sum rule  results indeed show that in the vacuum the $J^q_{\mu}$ and $J^s_{\mu}$ currents couple 
strongly to the $f_1(1285)$ and $f_1(1420)$, respectively, and that therefore 
these two states are mixed almost ideally.

\subsection{Finite density analysis of $f_1(1285)$}
Let us next turn to the main topic of this paper: the modification of the $f_1(1285)$ 
at finite density. 
This QCD sum rule analysis should provide a guideline for the expected mass shift of 
the $f_1(1285)$ in nuclear matter. 

\begin{figure}
\begin{center}
\vspace{-0.9cm}
\includegraphics[width=0.50\textwidth]{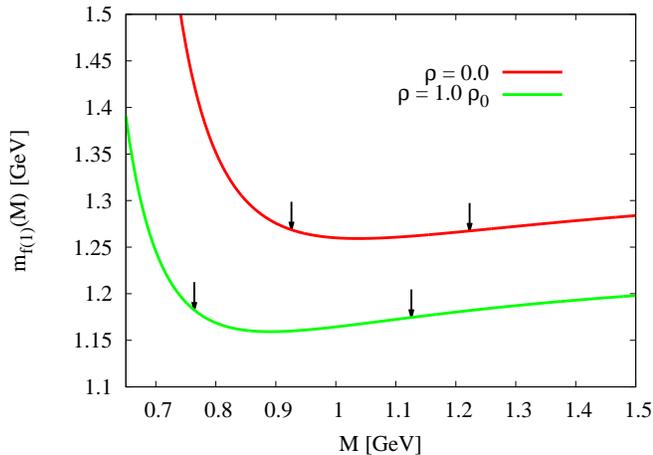}
\vspace{0.3cm}
\end{center}
\caption{The Borel curve for the mass in Eq.~(\ref{mass}) in the vacuum and at nuclear matter density.  
The arrows show the position of the minimum and maximum Borel mass.}
\label{borelwindow}
\end{figure}

The parameters used to quantify the density dependence of the 
various condensates are given in the lower part of Table~\ref{tab:parameters}. 
The corresponding result is shown in Fig.~(\ref{borelwindow}), where the Borel curves for 
both the vacuum and normal nuclear matter density $\rho_0$ are plotted. 
The flattest Borel 
curve in the latter case was obtained for a threshold parameter of  $\sqrt{s_0} = 1.49\,\mathrm{GeV}$. 

\begin{figure}
\begin{center}
\vspace{-0.8cm}
\includegraphics[width=0.50\textwidth]{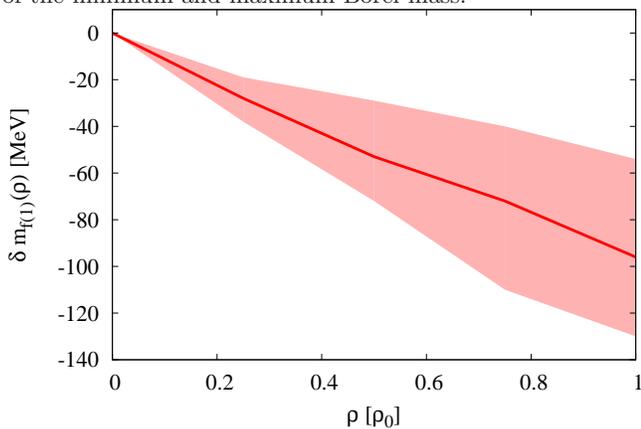}
\vspace{0.1cm}
\end{center}
\caption{The expected density dependence of the $f_1(1285)$ meson mass shift.  
The solid line is obtained with a value of 45 MeV for the $\pi N$ sigma term.  The lower and upper bounds were respectively 
obtained with $\sigma_{\pi N} = $ 60 MeV and 30 MeV . }
\label{mass}
\end{figure}

As it is seen in Table~\ref{tab:parameters}, the analysis was performed with a central value of 45 MeV 
for the $\pi N$ sigma term $\sigma_{\pi N}$. 
However, there are lattice results that show 
that $\sigma_{\pi N}$ might be smaller \cite{Durr:2015dna} while 
a recent phenomenological fit to experimental $\pi N$ scattering data suggests 
that it is bigger \cite{Hoferichter:2015dsa}. We have therefore performed the analysis for $\sigma_{\pi N} = $30 MeV and 60 MeV 
to check the sensitivity of our results on the $\pi N$ sigma term value, which 
is the largest source of error for the $f_1$ mass shift $\delta m_{f_1}$ and 
leads to the bands shown in Fig.~\ref{mass}, which depicts $\delta m_{f_1}$ as a function 
of density. Taking this uncertainty into account, we expect a mass shift of about $96 \pm 38$ MeV in nuclear medium.  

\section{Discussion and conclusion}

The mass shift in the sum rule is obtained by assuming a delta function pole for the $f_1(1285)$ in nuclear medium.  
It was however noted in an earlier work that for the vector meson sum rule, the changes of the OPE in the medium can 
also be satisfied with a smaller change in the mass and a simultaneous  increase in the width \cite{Leupold:1997dg}.  
It is likely that a similar effect also applies to the $f_1(1285)$ discussed in this paper. 
Therefore, our result should be considered as a maximum 
mass shift value expected at nuclear matter.  The experimental result for the $\omega$ 
suggests a small mass shift of -29 MeV and a larger increase in the width of 70 MeV \cite{Metag:2015lza}. 
Once the medium modification for the $f_1(1285)$ is experimentally observed, one can construct  QCD sum rules for the $\omega$ and $f_1(1285)$ separately and analyze how the changes in the corresponding masses and widths are related to the changes in the condensates.   From such analysis, one can then also estimate the effects of the factorizable part of the four quark condensate to the properties of the $\omega$ and $f_1(1285)$ meson. We leave such a detailed QCD sum rule analysis for both the $\omega$ and $f_1(1285)$ meson as future work.

The CLAS collaboration was able to clearly identify a sharp peak for the  $f_1(1285)$ on a proton 
target \cite{Dickson:2016gwc}. Performing the experiment on a nuclear target will involve several difficulties.   
First of all, the present mass shift is obtained with the $f_1(1285)$ meson at rest with respect to the nuclear medium.  
Experimentally selecting out low momentum $f_1(1285)$ will strongly suppress the signal.  
Moreover, reconstructing the $f_1(1285)$ from the hadronic final states will entail smearing and/or 
lost signal due to the rescattering of the final state with the medium.

However, as we have emphasized in this work, 
within the limit where disconnected diagrams can be neglected, 
such a measurement would be the first direct observation 
of a chiral symmetry restoration effect on hadron properties. 
Based on theoretical estimates on how much the chiral  order parameter would change 
at finite density, chiral symmetry is expected to be partially restored in nuclear matter. 
Experimentally observing the $f_1(1285)$ in nuclear matter would therefore serve 
as a test for these theoretical expectations and hence could 
shed light into the mechanism of how the mass of hadrons are generated. 
Considering the reward, the difficulties are worth overcoming.

\section*{Acknowledgments}
The work was supported by the Korea National
Research Foundation under the grant number
KRF-2011-0030621 and the Korean ministry of education under the grant number 2016R1D1A1B03930089.
The authors thank the Yukawa Institute for Theoretical Physics, Kyoto University, 
where this work was initiated  during the YITP-W-16-01 
``MIN16 - Meson in Nucleus 2016 -".   
T.K. was partially supported by a 
Grant-in-Aid for Scientific Research from the Ministry of Education,
Culture, Sports, Science and Technology (MEXT) of Japan
(Nos.16K05350,15H03663),
by the Yukawa International Program for Quark-Hadron Sciences.

\end{document}